\documentclass[aps,prl,amsmath,amssymb,superscriptaddress,showkeys,showpacs,twocolumn,floatfix]{revtex4-1}
\usepackage[final]{graphicx}
\usepackage{hyperref}
\usepackage{amsmath}
\usepackage{bbm}
\usepackage{amsfonts}
\usepackage{amssymb}
\usepackage{latexsym}
\usepackage{graphicx}
\usepackage[english]{babel}
\usepackage{multirow}
\usepackage{float}
\usepackage{url}
\usepackage{slashed}
\usepackage{xcolor}
\usepackage[utf8]{inputenc}
\usepackage{verbatim}
\usepackage{lipsum}

\usepackage[utf8]{inputenc}

\usepackage{diagbox}

\usepackage{mathtools}
\usepackage{amsfonts}
\usepackage{mathrsfs}
\usepackage{bbm}
\usepackage{slashed}

\usepackage{graphicx}
\usepackage{color}
\usepackage{array}
\usepackage{esint}
\usepackage{placeins}
\usepackage{booktabs}
\usepackage{makecell}
\usepackage{epstopdf}
\usepackage[caption=false]{subfig}

\usepackage{xspace}
\usepackage{siunitx}
\usepackage{hyperref}
\usepackage[nameinlink]{cleveref}
\usepackage{appendix}

\usepackage{xifthen}
\usepackage{xcolor}
\hypersetup{
	colorlinks,
	linkcolor={red!75!black},
	citecolor={blue!75!black},
	urlcolor={blue!75!black}
}
\usepackage{comment}

\newcommand{\be}{\begin{equation}}
\newcommand{\ee}{\end{equation}}
\newcommand{\ba}{\begin{array}}
\newcommand{\ea}{\end{array}}

\usepackage{ulem}
\newcommand{\beq}{\begin{eqnarray}}
\newcommand{\eeq}{\end{eqnarray}}

\begin{document}
\title{Numerical simulations of density perturbation and gravitational wave production from cosmological first-order phase transition}
\author{Jintao Zou}
\affiliation{Department of Physics and Chongqing Key Laboratory for Strongly Coupled Physics, Chongqing University, Chongqing 401331, P. R. China}

\author{Zhiqing Zhu}
\affiliation{University of Chinese Academy of Sciences (UCAS), Beijing 100049, P. R. China}
\affiliation{International Center for Theoretical Physics AsiaPacific, Beijing/Hangzhou 100190/310027, P. R. China}

\author{Zizhuo Zhao}
\affiliation{University of Chinese Academy of Sciences (UCAS), Beijing 100049, P. R. China}
\affiliation{International Center for Theoretical Physics AsiaPacific, Beijing/Hangzhou 100190/310027, P. R. China}

\author{Ligong Bian}\email{lgbycl@cqu.edu.cn}
\affiliation{Department of Physics and Chongqing Key Laboratory for Strongly Coupled Physics, Chongqing University, Chongqing 401331, P. R. China}



\begin{abstract}
We conducted three-dimensional lattice simulations to study the density perturbation and gravitational waves (GWs) during first-order phase transition (FOPT). We find that for phase transition strength $\alpha > 1$, the forward motion of bubble walls becomes the primary source, whereas for $\alpha < 1$, the dominant contribution to the density perturbation comes from the delay of vacuum decay. Additionally, the power spectrum of density perturbations generated by the phase transition exhibits a slope of $k^3$ at small wavenumbers and $k^{-1.5}$ at large wavenumbers. Furthermore, we calculated the GW power spectra, which exhibit the slope of $k^3$ at small wavenumbers and $k^{-2}$ at large wavenumbers. Our numerical simulations confirm that slow PTs can produce PBHs and provide predictions for the GW power spectrum, offering theoretical support for GW detection.

\end{abstract}	
\maketitle

\section{Introduction}
Primordial black holes (PBHs) are hypothesized to form in the early universe due to large density perturbations, offering a unique window into the physics of the early cosmos. There are currently many proposed mechanisms for the formation of PBHs, such as First-order phase transition (FOPT)~\cite{Hawking:1982ga,Crawford:1982yz,Kodama:1982sf,Khlopov:1998nm,Johnson:2011wt,Zou:2026wzi}, inflaton ultraslow-roll~\cite{Carr:1993aq,Ivanov:1994pa}, the collapse of cosmic strings~\cite{Hawking:1987bn,Caldwell:1995fu,Jenkins:2020ctp,Blanco-Pillado:2021klh}. Beyond these, scalar perturbations can also be induced by non-adiabatic fluctuations from various non-perturbative processes. Such as random distributions of PBHs~\cite{Liu:2021svg,Franciolini:2026fdv}, or the evolution of topological defects such as spherical domain walls~\cite{Vachaspati:2017hjw,Ferrer:2018uiu,Fu:2025qhf,Li:2025gld,Gelmini:2022uiu,Gelmini:2023aab,Gouttenoire:2023ftk,Gouttenoire:2023gbn,Bian:2022qbh} and solitons~\cite{Cotner:2018vug,Cotner:2016cvr,Li:2026yhv,Chen:2025oxo,Hou:2022jcd,Bai:2022kxq,Zhou:2024mea,Zhou:2015yfa,Wang:2023qxj}. Such mechanisms, along with scalar condensates~\cite{Dolgov:1992pu,Kasai:2002pu,Martin:2019nuw} and processes in a dissipative dark sector~\cite{Chang:2018bgx,Flores:2020drq,Domenech:2023afs,Chakraborty:2022mwu}, provide various pathways for generating large density perturbations. Unlike stellar-mass black holes formed from collapsing stars, PBHs can span a wide range of masses, from sub-planetary to stellar and even supermassive scales, depending on their formation mechanism and the underlying physics. They have attracted considerable attention as potential candidates for dark matter, sources of high-energy phenomena, and contributors to the gravitational wave background observed by detectors such as LIGO-Virgo~\cite{Sasaki:2016jop,Bird:2016dcv,Clesse:2016vqa,Hutsi:2020sol,Hall:2020daa,Franciolini:2021tla,He:2023yvl}, and the planned LISA~\cite{Armano:2016bkm,LISA:2017pwj,LISA:2022yao}, Taiji~\cite{Hu:2017mde,Ruan:2018tsw,TaijiScientific:2021qgx} as well as TianQin mission~\cite{TianQin:2015yph,Luo:2020bls,TianQin:2020hid}. 

In the early universe, FOPT proceeds through the nucleation, expansion, collision, and percolation of vacuum bubbles, and these bubbles release significant latent heat and can generate large density fluctuations from the random generation of vacuum bubbles~\cite{Liu:2021svg}. This topic has aroused sustained and growing interest because the FOPT is also an important source of stochastic gravitational wave (GW) background~\cite{Hindmarsh:2020hop,Bian:2025ifp}. During FOPT, the anisotropic stresses produced by vacuum bubble collisions~\cite{Kosowsky:1992rz,Kosowsky:1992vn,Kosowsky:1991ua,Child:2012qg,Cutting:2018tjt,Cutting:2020nla,Li:2023yzq,Zhao:2022cnn,Di:2020kbw}, sound waves~\cite{Giblin:2013kea,Hindmarsh:2015qta,Hindmarsh:2017gnf,Hindmarsh:2013xza,Cutting:2019zws,Cutting:2022zgd,Giblin:2014qia}, and magnetohydrodynamic turbulence (MHD)~\cite{Kosowsky:2001xp,Kamionkowski:1993fg,Neronov:2020qrl,RoperPol:2019wvy,Caprini:2009yp,Kahniashvili:2008pe,Kahniashvili:2009mf,Caprini:2015zlo,Wansleben:2019} generate GW radiations. The resulting GW spectrum encodes information about the properties of the phase transition, such as its strength, duration, and characteristic energy scale. Observing these GWs with current and future detectors provides a direct probe of the underlying particle physics responsible for the FOPT~\cite{
Athron:2023xlk,Caldwell:2022qsj,Bian:2021ini}, including extensions to the Standard Model, such as additional scalar fields or symmetry-breaking mechanisms.

In this work, we perform numerical simulations to investigate the density perturbation in the early universe, originating from the FOPT. We compare the effects of different phase transition rates and strengths on the density perturbation and the GW spectra generated by bubble collisions during the FOPT. Our findings indicate that the randomness of quantum tunneling during phase transitions can induce curvature perturbations, which can be orders of magnitude larger than primordial perturbations from inflation~\cite{Liu:2021svg}. Furthermore, smaller values of $\beta/H$ favor PBH formation. In our numerical simulations, the case with $\beta/H = 6$ successfully collapse into PBHs. For convenience, we adopt speed of light $c=8\pi G=1$ throughout this study.

\section{PBHs FROM PTs}
We consider a phase transition involving the scalar field $\phi$, and the associated potential can be expressed as~\cite{Cutting:2018tjt,Cutting:2020nla}
\begin{equation}
    V(\phi)=\frac{1}{2} M^2 \phi^2 + \frac{1}{3} \kappa \phi^3 + \frac{1}{4} \lambda \phi^4.
\end{equation}
During the transition, vacuum bubbles nucleate with an exponential rate expressed as~\cite{Coleman:1977py,Enqvist:1991xw}
\begin{equation}
    \Gamma(t) = \Gamma_0 e^{\beta t},
    \label{Eq: Vacuum bubbles nucleate}
\end{equation}
where $\Gamma_0$ is the initial nucleation rate, and $\beta$ is a key parameter that governs the rate’s increase over time. After nucleation, the field evolves according to the Klein-Gordon equation in an expanding universe~\cite{Child:2012qg}
\begin{equation}
    \ddot{\phi} + 3 H \dot{\phi} - \frac{\nabla^2 \phi}{a^2} +\frac{\partial V(\phi)}{\partial \phi} = 0,
\end{equation}
the spatially averaged energy density and pressure can be decomposed into four components~\cite{Adshead:2019lbr}
\begin{align}
    \rho_{tot} &= \rho_K + \rho_G + \rho_V + \rho_r \\
    p &= \rho_K - \frac{1}{3}\rho_G - \rho_V + \frac{1}{3}\rho_r,
\end{align}
where $\rho_K=\phi^{\prime 2}/{2a^2}$, $\rho_G=(\partial_i \phi)^2/{2a^2}$, $\rho_V$, and $\rho_r$ represent the spatially averaged kinetic energy density, gradient energy density, potential energy density, and energy density of background radiation, respectively. The PT strength is defined as $\alpha = \Delta V/\rho_r$, which represents the ratio of vacuum energy released to the background radiation energy density. The continuity equation can be written as
\begin{equation}
     \frac{{\mathrm{d} (\rho_K + \rho_G + \rho_r)}}{\mathrm{d} \tau} + 3\mathcal{H}(2\rho_K + \frac{2}{3}\rho_G + \frac{4}{3}\rho_r) 
    = -\frac{{\mathrm{d} \rho_V}}{\mathrm{d} \tau},
    \label{continuity equation}
\end{equation}
where, $\mathcal{H}=a^{\prime}/a$ is the conformal Hubble parameter related to the usual Hubble parameter via $\mathcal{H}=aH$ with $a^{\prime} \equiv \partial a / \partial \tau$, the conformal time $\tau$ is related to the cosmic time $t$ via $d\tau=dt/a$. The corresponding Friedmann’s equation is given as 
\begin{equation}
    \mathcal{H}^2 = \frac{8\pi}{3}a^2\rho_{tot}.
    \label{Friedmann equation}
\end{equation}

During the PT, asynchronous vacuum tunneling results in regions within each Hubble-sized volume where the false vacuum decay is delayed. In an expanding universe, while the radiation energy density decreases, the vacuum energy remains constant. This delay in false vacuum decay increases the local energy density in these regions. During the intermediate to late stages of a FOPT, density perturbations on the Hubble scale cross the horizon, leading to significant fluctuations with 
\begin{equation}
    \delta = \frac{\rho_{\rm{inside}}-\rho_{\rm{outside}}}{\rho_{\rm{outside}}},
\label{Eq:delta}
\end{equation}
where $\rho_{inside}$ and $\rho_{outside}$ represent the energy densities inside and outside the delayed decay region, respectively. If this overdensity exceeds the threshold $\delta_c$, the Hubble patch collapses into a PBH~\cite{Musco:2004ak,Harada:2013epa}. The numerical simulations suggest that the value of threshold is within the range of 0.40 to 0.67~\cite{Musco:2018rwt,Escriva:2019phb}. A recent study~\cite{Kierkla:2025vwp} points out a gauge inconsistency in certain works. Many studies on overdense regions in phase transitions use the flat gauge, while the threshold $\delta_c$ for PBH formation is typically calculated in the comoving gauge. This study suggests that the values of $\delta_c$ in these two different gauges can differ by an order of magnitude. However, it is worth noting that this work does not consider other contributions, especially the energy density in bubble walls may play an important role in the formation of PBHs during FOPT~\cite{Lewicki:2023ioy,Flores:2024lng,Hashino:2025fse}. Ref.~\cite{Carr:2026hot} suggests that this conclusion might change when the gradient energy of bubble walls is taken into account. Therefore, until this issue is fully resolved, we optimistically take $\delta_c = 0.45$~\cite{Jedamzik:1999am,Green:2004wb,Musco:2004ak,Musco:2012au,Harada:2015yda} and leave a more rigorous gauge-consistent analysis for future work. When the energy density in a Hubble volume exceeds this critical threshold, the entire mass within the Hubble horizon collapses to form a PBH, and its mass can be expressed as~\cite{Liu:2021svg,Goncalves:2024vkj}
\begin{equation}
    M_{\text{PBH}} \approx \gamma V_H \rho_c = 4\pi \gamma H^{-1}(t_{\text{PBH}}),
\label{PBH mass}
\end{equation}
where $\rho_c=3H^2(t_{\text{PBH}})$ is the critical density, $\gamma \leq 1$ is a correction factor associated with the specific physical processes of gravitational collapse, $t_{\text{PBH}}$ is the PBH formation time, the Hubble volume is $V_H=\frac{4}{3}\pi H(t)^{-3}$, and the fraction of PBHs from FOPT is~\cite{Hashino:2021qoq}
\begin{equation}
f_{\mathrm{PBH}} = \left(\frac{H(t_{\mathrm{PBH}})}{H(t_0)}\right)^2 \left(\frac{a(t_{\mathrm{PBH}})}{a(t_0)}\right)^3 P(t_n) \frac{1}{\Omega_{\mathrm{DM}}},
\label{PBH abundance}
\end{equation}
where $\Omega_{\mathrm{DM}}=0.245$ denotes the normalized ratio of the present dark matter density to the total energy density, $t_0$ is the time of the first bubble nucleation, $t_n$ is the time for bubbles starting to nucleate in those Hubble volumes with postponed vacuum decay, and $P(t_n)$ represents the probability that a Hubble volume collapses into a PBH at time $t_n$, with the following expression~\cite{Hashino:2021qoq,Liu:2021svg}
\begin{equation}
P(t_n)=\exp\left[-\frac{4\pi}{3}\int_{t_0}^{t_n}\frac{a^3(t)}{a^3(t_{\text{PBH}})}H^{-3}(t_{\text{PBH})}\Gamma(t)dt\right].
\label{P_t}
\end{equation}
This study relies entirely on numerical simulations, without such approximations. Next, we define the variance of density perturbations as
\begin{equation}
    \sigma^2_\delta(k=l^{-1},t) =\langle \delta_l^2(k,t) \rangle - \langle \delta_l(k,t) \rangle^2,
\label{Eq:sigma}
\end{equation}
where $l$ is the characteristic spatial scale. It is worth noting that, here, $\langle \delta_l(k,t) \rangle=0$. The power spectrum of these perturbations plays a crucial role in understanding the characteristics of density fluctuations. The relationship between the dimensionless power spectrum of density perturbations and the variance of density perturbations is as follows
\begin{align}
    &\sigma^2_\delta(k=l^{-1},t) = \int_0^{1/l} \mathcal{P}_\delta(k,t) \frac{dk}{k},\\
    &\mathcal{P}_\delta(k,t) = \frac{k^3}{2\pi^2}|\tilde{\delta}_l(k,t)|^2.
    \label{Eq:P_delta}
\end{align}
In our simulation, we divide the simulation region into multiple blocks. For the density perturbations involved in the density perturbation power spectrum, unlike the previous Eq.~(\ref{Eq:delta}), we treat each block as a point, which means $\delta=(\rho-\bar{\rho})/\bar{\rho}$. The relationship between the density perturbation power spectrum and the curvature perturbation power spectrum is given by~\cite{Josan:2009qn,Ando:2018qdb}
\begin{equation}
    \mathcal{P}_\delta(k, t)=\frac{16}{3}\left(\frac{k}{a H}\right)^2 j_1^2(k / \sqrt{3} a H) \mathcal{P}_{\mathcal{R}}(k),
\label{Eq:delta_R1}
\end{equation}
where $j_1$ is a spherical Bessel function, whose value at horizon ($k_H=aH$, $\tau_H$ is the time of entry into the Hubble horizon) crossing simply reads
\begin{equation}
    \mathcal{P}_\mathcal{R}(k_H)=\frac{3\mathcal{P}_\delta(k_H,\tau_H)}{16j^2_1(1/\sqrt{3})}.
\label{Eq:delta_R}
\end{equation}

During a FOPT, vacuum bubbles are generated throughout space via thermal tunneling and then rapidly expand. As more and more bubbles form and collide, the initially uniform spatial structure is disrupted, resulting in an asymmetric energy distribution. The walls of these bubbles carry high energy, and when they collide, this energy is intensely released, creating asymmetric disturbances. This asymmetry breaks the spherical symmetry of the system, causing violent oscillations in spacetime curvature. These oscillations propagate as waves throughout the universe, producing GWs. 

To investigate the production of GWs, we concentrate on the tensor sector of the FLRW metric perturbations
\begin{equation}
    ds^2 = a(\tau)^2 \left[ -d\tau^2 + (\delta_{ij} + h_{ij}) dx^i dx^j \right],
\end{equation}
where $h_{ij}$ represents the small perturbation of the metric. GWs only require considering the transverse and traceless components ($\partial^i h_{ij} = 0$, $h_{ii} = 0$). The linearized Einstein equations yield the equation of motion for $h_{ij}$
\begin{equation}
    h''_{ij} - \partial_k \partial^k h_{ij} + 2\mathcal{H}h'_{ij} - 2(2\mathcal{H}' + \mathcal{H}^2)h_{ij} = 16\pi G T_{ij}^{\mathrm{TT}},
\end{equation}
where $T_{ij}^{\mathrm{TT}}$ denotes the transverse-traceless component of the stress-energy tensor. Therefore, the energy-momentum tensor and the energy density for GWs are given by~\cite{Hindmarsh:2015qta}
\begin{align}
T_{\mu\nu}^{\text{GW}} &= \frac{1}{32 \pi G} \langle \partial_\mu h_{ij} \partial_\nu h_{ij} \rangle, \\
\rho_{gw} &= \frac{1}{32\pi G} \langle \dot{h}_{ij}^2 \rangle,
\end{align}
the GW power spectrum is then
\begin{equation}
    \Omega_{gw}(t) = \frac{1}{24\pi^2\mathcal{H}^2V}\sum_{i,j}\int\mathrm{d}\Omega\left|\mathbf{k}\right|^3\left| h_{ij}'(k,t) \right|^2,
\end{equation}
where $V$ is the simulation volume. 

\section{Numerical results}
\label{NUMERICAL RESULTS}
In this simulation, we employ a lattice of $L^3=(512dx)^3$ to model a region with $LH=4$. Corresponding to a total of $4^3$ Hubble volumes. To calculate the density perturbations defined in Eq.~(\ref{Eq:delta}), it is essential to identify regions where the false vacuum decay is delayed. During the FOPT, thermal tunneling leads to local energy density variations. In our simulation, we monitor the 64 Hubble volumes and record the specific time when the first bubble center nucleates within each volume. We then rank these volumes chronologically based on their nucleation times. Specifically, the last 16 volumes to undergo nucleation (representing the final 25\% of the sample) are defined as postponed Hubble volumes. This criterion allows us to isolate regions with the most significant local energy density excess relative to the average background. In slow transitions, these postponed regions naturally correspond to areas that remain in the false vacuum state longer due to the low nucleation rate.

We simulate the thermal tunneling process in the false vacuum through random bubble nucleation. For the vacuum bubble nucleation probability given by Eq.~(\ref{Eq: Vacuum bubbles nucleate}), in this work, we explore a wide range of parameters, specifically $\alpha=$ 0.5, 1, 5, 10 and $\beta/H=$ 6, 8, 10, 12. For the potential parameters, we choose $M=0.002$, $\kappa=-0.04$, and $\lambda=7$ in order that the successful completion of the FOPT with the initial bubble radius and wall thickness being calculated as $12dx$ and $4dx$ through the {\it findbounce}~\cite{Guada:2020xnz}, respectively. The mean bubble separation is obtained as $R_*=(L^3/N_b)^{\frac{1}{3}}$ with $N_b$ being the number of the generated bubbles during the FOPT processes. While if these PT parametere can be achieved in the realistic particle physics models is still in debate, see Ref.~\cite{Goncalves:2024vkj,Huang:2025hos} for successful realization in the real and complex singlet extension of the Standard Model and challenges in classically conformal gauge-Higgs theories~\cite{Kierkla:2025vwp}, we adopt these values to provide a model-independent study. Our results focus on establishing the relationship between these phase transition parameters and the resulting density perturbations, providing a benchmark for various potential early-universe scenarios. The spatial and temporal lattice spacings are defined as $\mathrm{d}x=L/N$ and $\mathrm{d}t=\mathrm{d}x/5$, respectively. At the end simulation, the bubble wall thickness falls within the range of $1.2dx$ to $1.6dx$, with the contributions from smaller scales ($\log_{10}[kR_*]>1.3$) to the density fluctuation and GW spectra being dropped. The numerical calculations are conducted with a code developed based on {\it pystella}~\cite{Weiner:2021zj}.
\begin{figure*}[t!]
    \centering
    \includegraphics[width=0.8\textwidth]{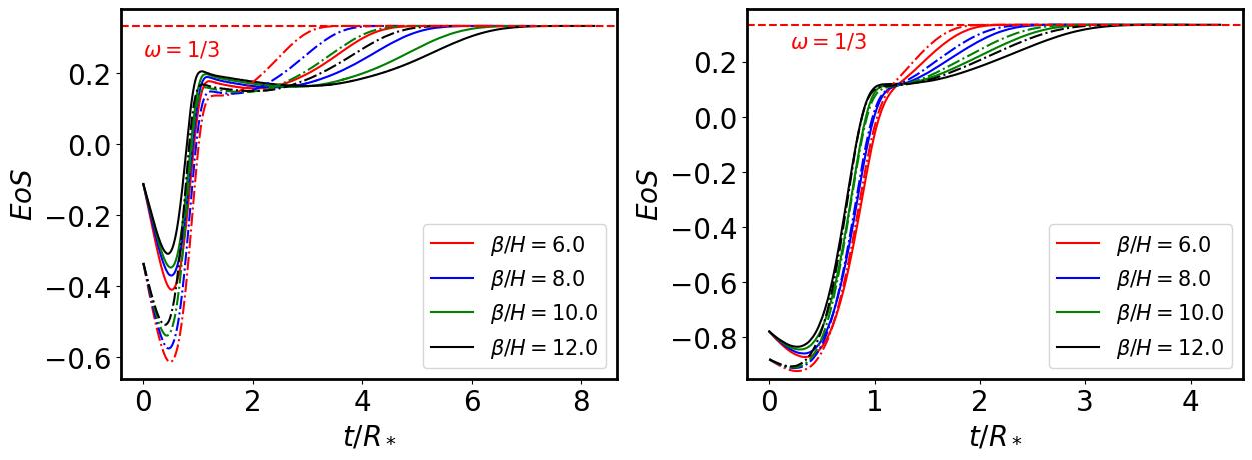}
    \caption{Equation of state (EOS) evolution for weak phase transition strength (left, solid line $\alpha$=0.5, dash-dotted line $\alpha$=1) and strong phase transition strength (right, solid line $\alpha$=5, dash-dotted line $\alpha$=10). Four colors represent different $\beta/H$ values. In all cases, $\omega$ transitions from $\omega \to -1$ (vacuum energy domination) to $\omega = 1/3$. A smaller $\alpha$ results in a longer transition time to reach $\omega = 1/3$.}
    \label{EoS}
\end{figure*}

The evolution of the average energy density of each component follows the evolution described by continuity equation Eq.~(\ref{continuity equation}) and the Friedmann equation Eq.~(\ref{Friedmann equation}). As the phase transition progresses, the vacuum bubbles expand, converting the vacuum energy $\rho_V$ into the bubble wall energy $\rho_w$ (including the kinetic and gradient energy of the wall) and the background radiation energy. For the cosmic FOPT, the numerical simulation can be performed with the scalar-fluid system, therein the vacuum energy would released to the fluid motion and one can obtain the sound wave contribution to the GWs~\cite{Cutting:2019zws,Hindmarsh:2017gnf,Hindmarsh:2015qta,Hindmarsh:2013xza}. At the end of the PT, one have the background radiation energy density dominate the energy density, see Fig.~\ref{Fig:rho} in {\it Supplemental Material} for details. The Equation of State (EoS), which is computed by $\omega=p/\rho_{tot}$, is shown in Fig.~\ref{EoS}. It can be seen that in the beginning, the universe was dominated by vacuum energy, also known as dark energy, with a value of $\omega \to -1$. As the vacuum energy is continuously converted into kinetic and gradient energies, $\omega$ continues to increase. After the phase transition of the entire space is completed, the universe ultimately admit $\omega=1/3$.
\begin{figure}[t!]
    \centering
    \includegraphics[width=0.45\textwidth]{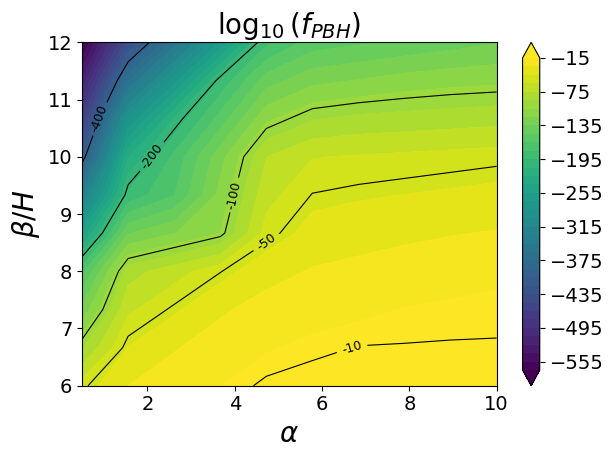}
    \caption{The PBH abundance variation with respect to the PT strength parameter $\alpha$ and the inverse duration $\beta/H$.}
    \label{fig:PBH abundance}
\end{figure}

By applying Eqs.~(\ref{PBH abundance}) and (\ref{P_t}), here, $H$ and $a$ are dynamically evolving parameters during the simulation process. We calculated the PBH abundance for different values of $\alpha$ and $\beta/H$, and the results are presented in Fig.~\ref{fig:PBH abundance}. On the whole, we find that the PBH abundance increases with $\alpha$ and decreases with $\beta/H$, which is consistent with the conclusions of Refs.~\cite{Liu:2021svg,Kawana:2022olo,Cai:2024nln}. Notably, this dependence is stronger on $\beta/H$ and weaker on $\alpha$. In particular, when $\alpha \gtrsim 10$, the influence of $\alpha$ on the PBH abundance is further suppressed. Furthermore, the results indicate that PBH formation occurs for slow phase transitions with $\beta/H = 6$, see {\it Supplemental Material} for more details during the PT processes. This is consistent with the conclusion of Refs.~\cite{Gouttenoire:2023naa,Gouttenoire:2023bqy,Gouttenoire:2023pxh} that PBHs are more likely to form when $\beta/H<6$.
\begin{figure*}[t!]
    \centering
    \includegraphics[width=0.8\textwidth]{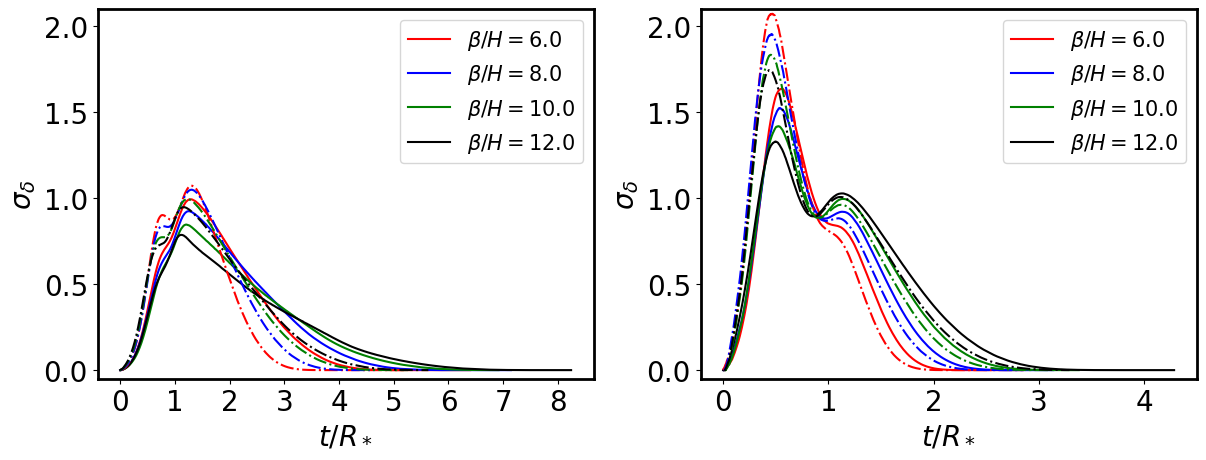}
    \caption{Evolution of $\sigma_\delta$ at the Hubble scale for weak phase transition strength (left, solid line $\alpha$=0.5, dash-dotted line $\alpha$=1) and strong phase transition strength (right, solid line $\alpha$=5, dash-dotted line $\alpha$=10).}
\label{Fig:sigma_all}
\end{figure*}

\begin{figure}[htbp]
    \centering
    \includegraphics[width=0.45\textwidth]{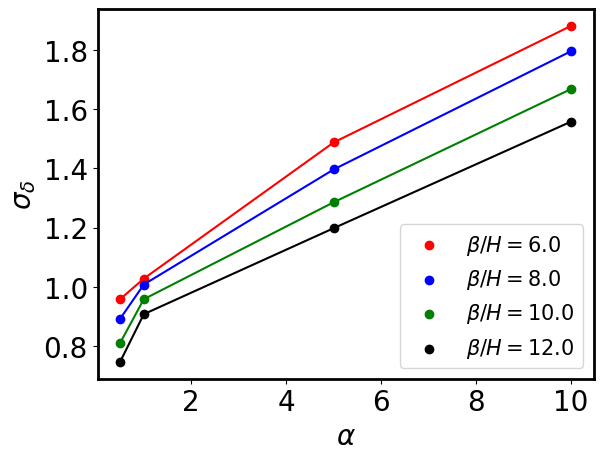} 
    \caption{
    Variation of the standard deviation $\sigma_\delta$ of accumulated overdensity perturbations with respect to the PT parameters $\alpha$ and $\beta/H$.}
    \label{sigma_apha}
\end{figure}

For the density perturbation, we compute the value of $\sigma_\delta$ at the Hubble scale using Eq.~(\ref{Eq:sigma}), and the corresponding evolution is presented in Fig.~\ref{Fig:sigma_all}. Where, the $\sigma_\delta$ for PTs of $\alpha=0.5, 1$ tends to reach the maximum when $t/R_*\sim 1.2$ mostly come from the energy transformation in the simulation volume, which is influenced by the forward motion of bubble walls. Differently, the $\sigma_\delta$ reach maximum at around $t/R_*\sim 0.5$ for strong PTs scenarios with $\alpha$=5, 10 sourced from the delay of vacuum decay in different Hubble volumes. This observation is reflected in Fig.~\ref{bubbleslice} and Fig.~\ref{Fig:rho} of the {\it Supplemental Material}, where we present details on the PT process and the energy density evolution. We plot the maximum values of $\sigma_\delta$ as a function of $\alpha$ for different $\beta/H$ in Fig.~\ref{sigma_apha}, which indicates that the magnitude of $\sigma_\delta$ grows as the PT strength increases.

\begin{figure*}[t!]
    \centering
    \includegraphics[width=0.8\textwidth]{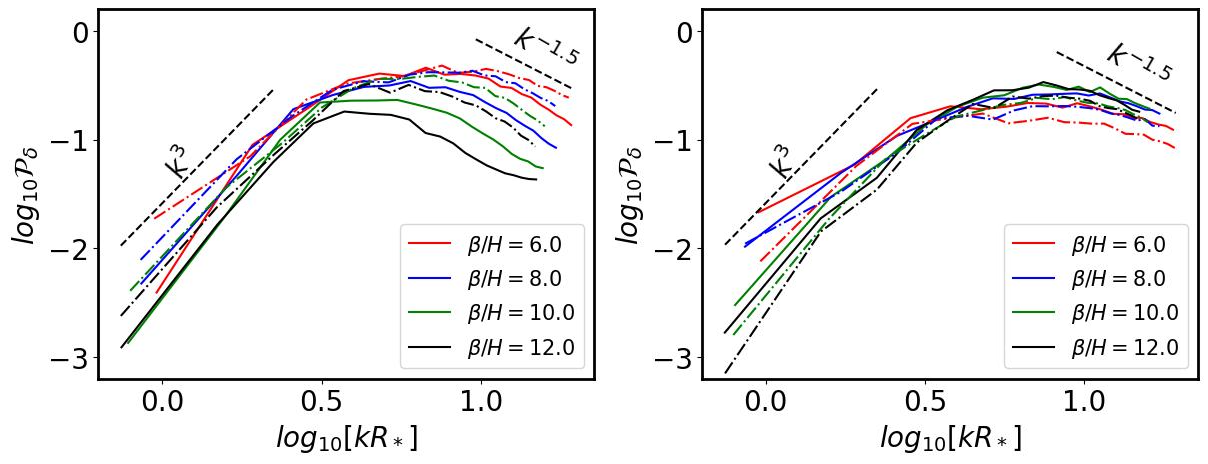}
    \caption{The power spectra of density perturbations for different $\beta/H$ values during the late stage of the phase transition. Left panel: solid line $\alpha$=0.5, dash-dotted line $\alpha$=1; Right panel: solid line $\alpha$=5, dash-dotted line $\alpha$=10.}
\label{Fig:delta_sp}
\end{figure*}

Fig.~\ref{Fig:delta_sp} shows the power spectrum of density perturbations at the completion of the phase transition. For $log_{10}[kR_*]\textless 0.5$, the slope follows $k^3$, while for $0.8 \textless log_{10}[kR_*]\textless 1.3$, the slope is $k^{-1.5}$. 
In the infrared range ($k<k_H$), the vacuum decay process has little effect and can be treated as uncorrelated. Therefore, causality requires that, for small wavenumbers, the perturbation $\delta_l(k,t) \propto k^0$~\cite{Liu:2022lvz}, This behavior is similar to the perturbations caused by a Poisson distribution of primordial black holes~\cite{Papanikolaou:2020qtd,Afshordi:2003zb}. In the ultraviolet range ($k>k_H$), the slope of the power spectrum is primarily determined by small-scale regions where the energy density changes abruptly. During the evolution, the vacuum energy difference across the bubble wall can be treated as Heaviside step function, whose Fourier amplitude scales as $\delta_l(k) \propto k^{-2}$ for a three-dimensional spherically symmetric distribution. Consequently, the dimensionless power spectrum should follow $\mathcal{P}_\delta(k) \propto k^3 \cdot (k^{-2})^2 = k^{-1}$ (The analytical derivation is provided in the {\it Supplementary Material}). However, in numerical simulations, the presence of kinetic and gradient energies smoothens the energy transition across the wall, resulting in a final slope of $-1.5$ for the dimensionless power spectrum. Notably, no clear correlation is observed between the power spectrum slope and the parameters $\beta/H$ or $\alpha$. When the curvature perturbation at the Hubble horizon scale can be derived from Eq.~(\ref{Eq:delta_R}), which can be quantitatively characterized by the following fitting equation~\cite{Liu:2022lvz}
\begin{equation}
    \mathcal{P}_\mathcal{R}(\alpha,\beta/H) = A \alpha^2[f(\beta/H)]^2(kR_H)^3,
\label{Eq:Fit PR}
\end{equation}
where the $f(\beta/H) \equiv \sigma_\delta/\alpha$. The fitting result is $A=6.10$, compared to the value 34.5 in Ref.~\cite{Liu:2022lvz} for $\alpha\lesssim 1$ scenario.
\begin{figure*}[t!]
    \centering
    \includegraphics[width=0.85\textwidth]{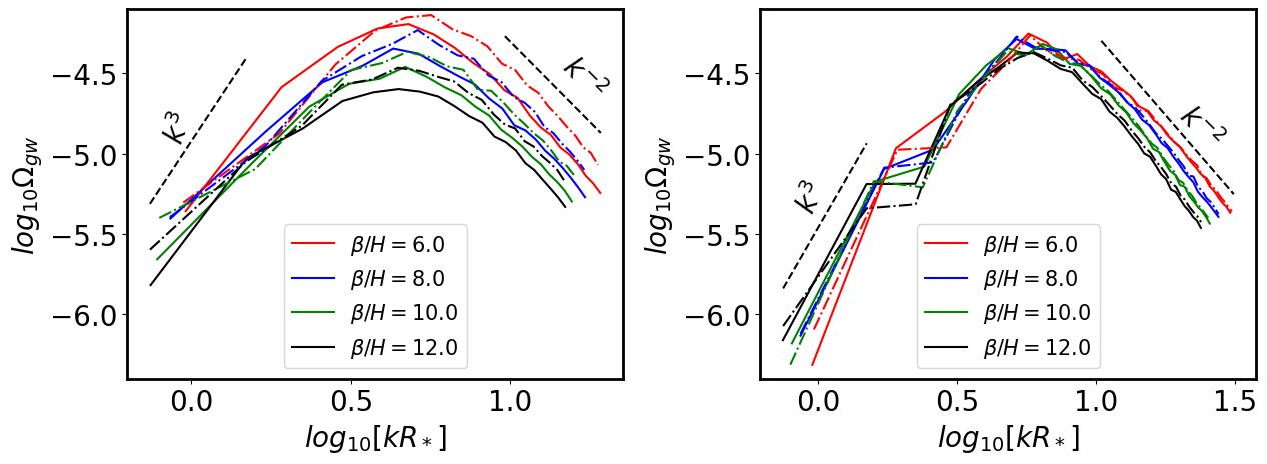}
    \caption{The energy density spectra of GWs for different $\beta/H$ values at the late stage of the phase transition. Left panel: solid line $\alpha$=0.5, dash-dotted line $\alpha$=1; Right panel: solid line $\alpha$=5, dash-dotted line $\alpha$=10.}
    \label{gw}
\end{figure*}

When vacuum bubbles expand and collide, the large amount of energy carried by their walls is released violently and creates asymmetric disturbances, making it a source of GWs~\cite{Falkowski:2012fb,Konstandin:2011ds}. With the proceeding of PTs, the magnitudes of the GW spectra grow, as show in the {\it Supplemental Material}. In Fig.~\ref{gw}, we show the energy density spectra of GWs for FOPTs. For $\log_{10}[kR_*]\textless 0.5$, the GW power spectrum exhibits a steep rise with a slope of $k^3$, which is a little bit different for the $\alpha=0.5$ case and the discrepancy may be caused by the numerical fault. In the range of $0.8 \textless \log_{10}[kR_*]\textless 1.3$, the slope is $k^{-2}$, Moreover, our analysis reveals that the overall shape and amplitude of the power spectrum show little sensitivity to the parameters $\alpha$ and $\beta/H$. Meanwhile, according to our calculations for the $\alpha$ is $0.5$ and $1$, the magnitude is larger when $\beta/H$ is smaller, though this trend weakens for $\alpha = 5$ and $10$.

We perform a fit to the GW spectrum in the presence of cosmic expansion, the fitting function is given below~\cite{Zhao:2022cnn}
\begin{figure}[t!]
    \centering
    \includegraphics[width=0.45\textwidth]{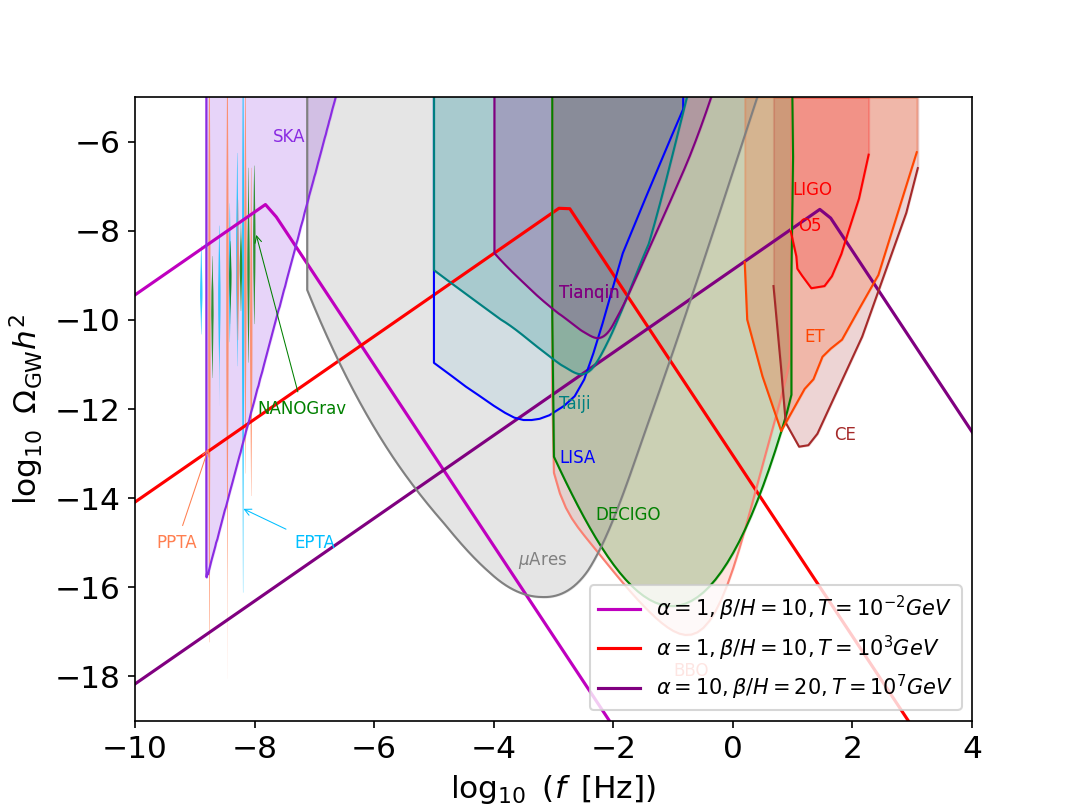}
    \caption{GW spectra for three scenarios: ($\alpha=1,\beta/H=10,T=10^{-2}$~GeV), ($\alpha=1,\beta/H=10,T=10^3$~GeV), and ($\alpha=10,\beta/H=20,T=10^7$~GeV). The sensitivity curves including EPTA~\cite{EPTA:2015qep}, PPTA~\cite{Shannon:2015ect}, NANOGrav~\cite{NANOGRAV:2018hou}, $\mu Ares$~\cite{Sesana:2019vho}, SKA~\cite{Carilli:2004nx}, LISA~\cite{LISA:2017pwj}, Taiji~\cite{Ruan:2018tsw}, DECIGO~\cite{Kawamura:2011zz}, LIGO~\cite{LIGOScientific:2014pky}, CE~\cite{Reitze:2019iox}, and ET~\cite{Punturo:2010zz}.}
    \label{gw_fit}
\end{figure}
\begin{equation}
\Omega_{\text{gw}} = \tilde{\Omega}(k_p) \frac{(a + b)^c k_p^b k^a}{\left( bk_p^{(a + b)/c} + ak^{(a + b)/c}\right)^c},
\label{Eq:Fit GW}
\end{equation}
we obtain $a=0.93,b=2.03,c=0.03$ from the fit. We adopt the parametrization of the GW spectrum given in Ref.~\cite{Cutting:2018tjt,Zhao:2022cnn}
\begin{equation}
h^{2} \Omega_{\mathrm{GW}}(f) = h^{2} \tilde{\Omega}(f_{p}) \frac{(a+b)^{c} f_{p}^{b} f^{a}}{\left(b f_{p}^{(a+b)/c}+a f^{(a+b)/c}\right)^{c}},
\label{Eq:GW_fit}
\end{equation}
the magnitude of the low‑frequency GW spectrum is $\tilde{\Omega}(f_{p}) = 1.67\times10^{-5} \left(\frac{100}{g_{\ast}(T_{\ast})}\right)^{1/3} \left(\frac{\alpha}{1+\alpha}\right)^{2} \left(\frac{H_{\ast}}{\beta}\right)^{2}$, with its low‑frequency peak located at $f_{p} = 1.6\times10^{-5} \frac{\beta}{H_{\ast}} \frac{T_{\ast}}{100} \left(\frac{g_{\ast}}{100}\right)^{1/6} \ \mathrm{Hz}$. In Fig.~\ref{gw_fit}, we present the FOPT GW spectra obtained by substituting the fitted parameters into Eq.~(\ref{Eq:GW_fit}). The results also show cases detectable by LIGO~\cite{LIGOScientific:2014pky}, Taiji~\cite{Ruan:2018tsw}, and SKA~\cite{Carilli:2004nx}.

\section{Conclusion and discussion}
In this study, we performed three-dimensional numerical simulations to investigate the PBH abundance, density perturbation, and GWs production during FOPTs. Our simulations show that regions undergoing delayed vacuum decay can indeed develop local overdensities, and that smaller values of $\beta/H$ facilitate the formation of such overdense regions. Additionally, we computed the abundance of PBHs and found that it increases with $\alpha$ but decreases with $\beta/H$, which is consistent with the conclusions reported in Refs.~\cite{Liu:2021svg,Kawana:2022olo,Cai:2024nln}. Notably, the result exhibits stronger sensitivity to variations in $\beta/H$ and a relatively weaker dependence on $\alpha$. In particular, when $\alpha \gtrsim 10$, the influence of $\alpha$ on the PBH abundance becomes further suppressed.

In the power spectrum of density perturbations at the end of the phase transition, the slope follows $k^3$ for $\log_{10}[kR_*] \textless 0.5$, while in the range $0.8 < \log_{10}[kR_*] \textless 1.3$, the slope is $k^{-1.5}$. We observed two peaks in the density perturbation. The first peak is caused by the delay of vacuum decay in different Hubble volumes, while the second peak originates from the moving forward behavior of the bubble walls. Moreover, for $\alpha < 1$, the peak due to the delay of vacuum decay dominates, whereas for $\alpha > 1$, the peak arising from the moving forward behavior of the bubble walls becomes dominant. We further note that when the fluid motion is considered the result for $\alpha > 1$ might deminished to some extent, since therein bubble wall energy would converted to the fluid energy and yield sound wave contribution to the GWs.

Since FOPT can generate significant curvature perturbations on small scales, the upper bound constraints on the power spectrum of curvature perturbations $\mathcal{P}_\mathcal{R}$ can be translated into limitations for the phase transition parameters $\alpha$ and $\beta/H$. Using the fitted curvature perturbation Eq.~(\ref{Eq:Fit PR}), our results yield less stringent constraints on $\alpha$ and $\beta/H$ compared to those reported in Ref.~\cite{Liu:2022lvz}, but this result is slightly more stronger than that reported in Ref.~\cite{Cai:2024nln}. Such a PT is found to be in tension with the Hubble tension explanation~\cite{Elor:2023xbz}. For more detailed on the BBN and $\Delta N_{eff}$ constraints on the PT parameters, we refer to Refs.~\cite{Bai:2021ibt,Deng:2023twb}. 

Furthermore, the GW spectra were shown to have the slope following $k^3$ and $k^{-2}$ at small and large wave numbers. We did not observe a significant dependence on $\alpha$ and $\beta/H$ for these slopes. Additionally, when analyzing the GW power spectrum at different stages of the phase transition, we did not find any notable changes in its slope. In the simulations with $\alpha = 0.5$ and $\alpha = 1$, we found that as $\beta/H$ decreases, the magnitude is larger. This finding is consistent with the conclusions of Ref.~\cite{Lewicki:2024ghw}. The final formation of PBH would require fully relativistic numerical simulations, the present work still provides reliable results on density perturbations from a FOPT.

\begin{acknowledgments}
We thank Jing Liu for fruitful discussion. This work is supported by the National Key Research and Development Program of China under Grant No. 2021YFC2203004, and by the National Natural Science Foundation of China (NSFC) under Grants Nos. 12322505, 12547101. We also acknowledges Chongqing Talents: Exceptional Young Talents Project No. cstc2024ycjh-bgzxm0020 and Chongqing Natural Science Foundation under Grant No. CSTB2024NSCQ-JQX0022.
\end{acknowledgments}

\bibliography{ref}

\clearpage

\onecolumngrid
\begin{center}
  \textbf{\large Supplemental Material}\\[.2cm]
\end{center}

In this supplemental material, we provide details on density perturbations and GW productions during the PT process.

\section{Phase transition process}
\begin{figure}[H]
    \centering
    \includegraphics[width=1\textwidth]{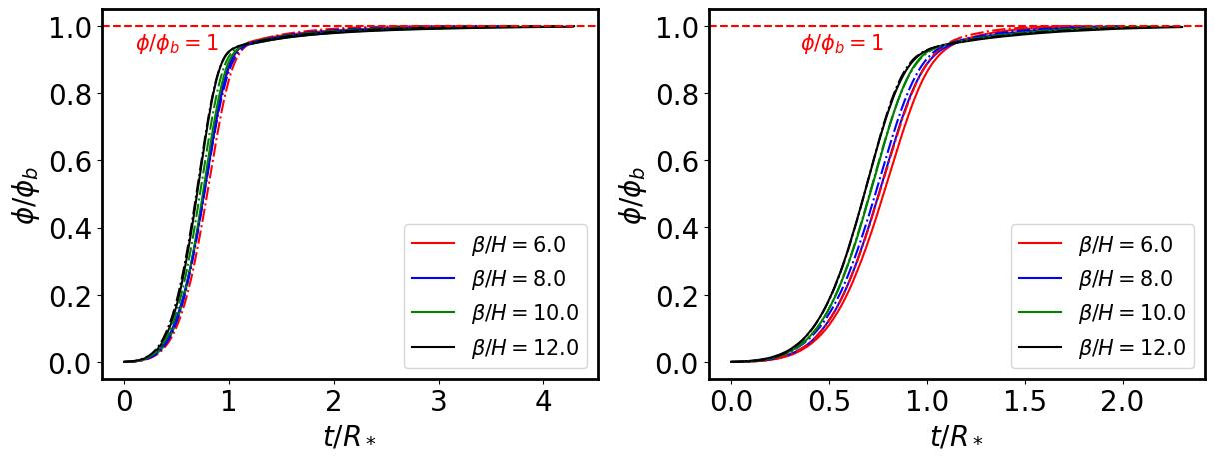}
    \caption{The mean field evolution under weak phase transition strength (left, solid line $\alpha$=0.5, dash-dotted line $\alpha$=1) and strong phase transition strength (right, solid line $\alpha$=5, dash-dotted line $\alpha$=10). Four colors represent different $\beta/H$.}
    \label{phievo}
\end{figure}

The evolution of the scalar field $\phi$ is depicted in Fig.~\ref{phievo}. It can be observed that the mean value of the field gradually increases from zero, corresponding to the nucleation and expansion of vacuum bubbles in space. Initially, the growth rate of the field’s mean value accelerates before slowing down and eventually reaching a plateau. Moreover, as $\alpha$ increases and $\beta/H$ becomes larger, the completion time of the phase transition, characterized by $t/R_*$, decreases. This is expected, as a stronger phase transition facilitates a more rapid completion of the process. The time of phase transition completion and the corresponding number of bubbles are presented in Tab.~\ref{tab:wide_table}.

\begin{table}[H]
    \caption{Bubble parameters for different values of $\beta/H$ and $\alpha$. Each cell contains two quantities: the bubble number $N_b$ of the generated bubbles during the FOPT processes, the time $t/R_*$ required to reach $\phi/\phi_b$=0.999.}
    \centering
    \renewcommand{\arraystretch}{1.2}  
    \setlength{\tabcolsep}{4pt}  
    \begin{tabular}{c|cccc}
        \hline
        $\alpha \backslash \beta/H$ &       6       &       8       &       10      &      12       \\
        \hline
        $ 0.5 $                     & $(281, 3.57)$ & $(391, 4.12)$ & $(521, 4.67)$ & $(601, 5.02)$ \\
        $  1  $                     & $(296, 2.71)$ & $(397, 3.15)$ & $(498, 3.56)$ & $(610, 3.93)$ \\
        $  5  $                     & $(289, 1.83)$ & $(388, 2.13)$ & $(481, 2.42)$ & $(601, 2.72)$ \\
        $ 10  $                     & $(278, 1.70)$ & $(387, 2.01)$ & $(500, 2.30)$ & $(598, 2.54)$ \\
        \hline
    \end{tabular}
    \label{tab:wide_table}
\end{table}

\begin{figure*}[!t]
    \centering
    \includegraphics[width=1\textwidth]{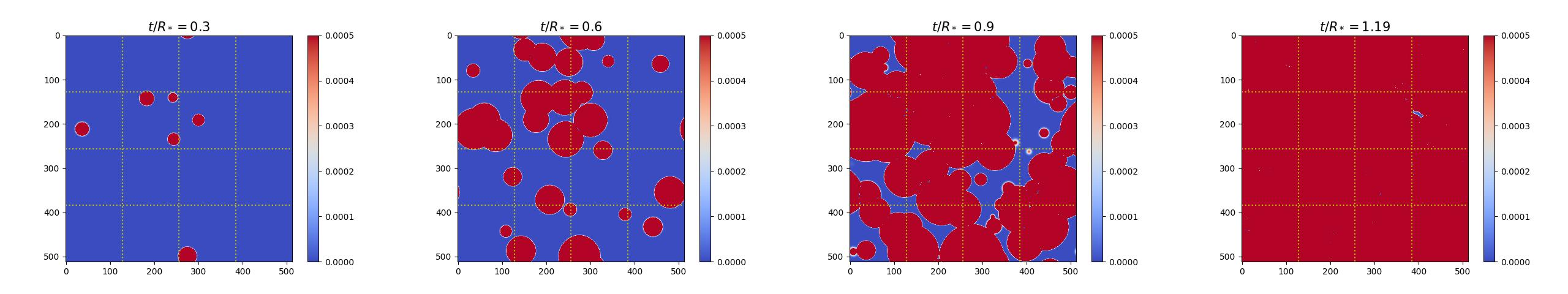}
    \caption{The bubble slice diagram for $\beta/H$ = 6 and $\alpha$ = 5. The dashed lines are presented to divide the simulation box to 64 Hubble volumes. }
    \label{bubbleslice}
\end{figure*}

\begin{figure*}[!t]
    \centering
    \includegraphics[width=1\textwidth]{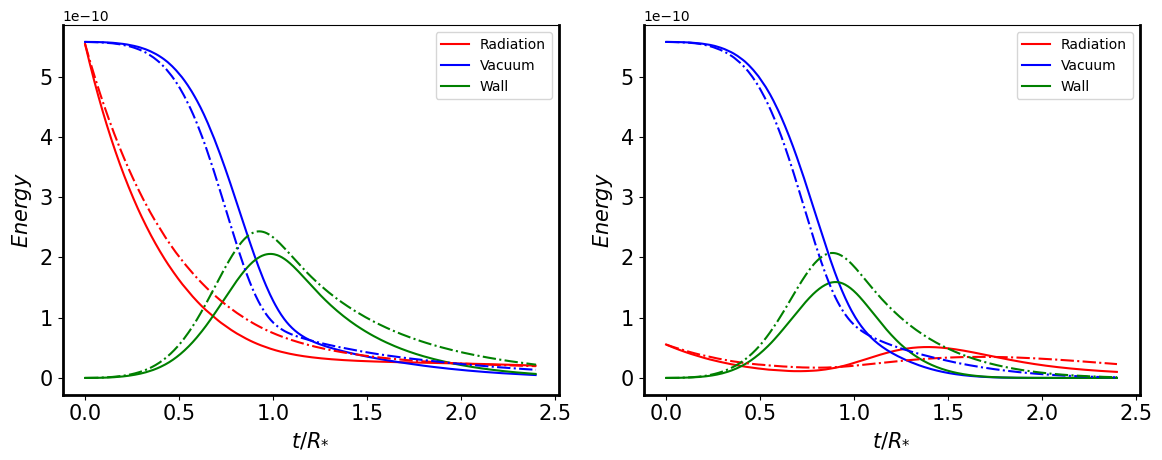}
    \caption{The energy density $\rho$ under weak phase transition strength (left, $\alpha = 1$, solid line $\beta/H=6$, dash-dotted line $\beta/H=10$) and strong phase transition strength (right, $\alpha = 10$ solid line $\beta/H=6$, dash-dotted line $\beta/H=10$).}
    \label{Fig:rho}
\end{figure*}

As an illustration in Fig.~\ref{bubbleslice}, we present bubble slices with $\beta/H=6$ and $\alpha=5$, which visually demonstrate the process of bubble expansion and phase transition. Clearly, one could have more energy transformation from bubble walls' moving forward with the proceeding of the PT. Fig.~\ref{Fig:rho} show the energy evolution of different components. During the FOPT, vacuum energy is released and converted into bubble wall energy and the background radiation energy. For a weak transition ($\alpha=1$), the initial radiation energy is large, and its evolution is dominated by cosmic expansion, resulting in a continuous decay. The bubble wall energy peaks at $t/R_\star \approx 1$ and subsequently decreases due to the expansion. In the strong transition case ($\alpha=10$), however, the radiation energy increases for a period after $t/R_\star \approx 1$ resulting from the energy released from the false vacuum. Additionally, a larger $\beta/H$ accelerates the entire process, leading to a higher and earlier peak in bubble wall energy, as well as a faster depletion of vacuum energy. For the method of scalar-fluid system simulation as in Ref.~\cite{Hindmarsh:2013xza,Hindmarsh:2015qta,Hindmarsh:2017gnf,Hindmarsh:2020hop}, we expect the bubble wall energy initially increases, reaching a peak around $t/R_\star \approx 1$, after which it starts to decrease. Indeed, as bubbles collide and merge, the energy of the bubble walls gradually converts into the background radiation energy. For this situation, the peak of the $\sigma_\delta$ from the forward motions of bubble walls might be deminished. However, the one resulting from the delay of vacuum decay in different Hubble volume still there.

\section{Evolution of the density perturbation}
\begin{figure*}[!t]
    \centering
    \includegraphics[width=1\textwidth]{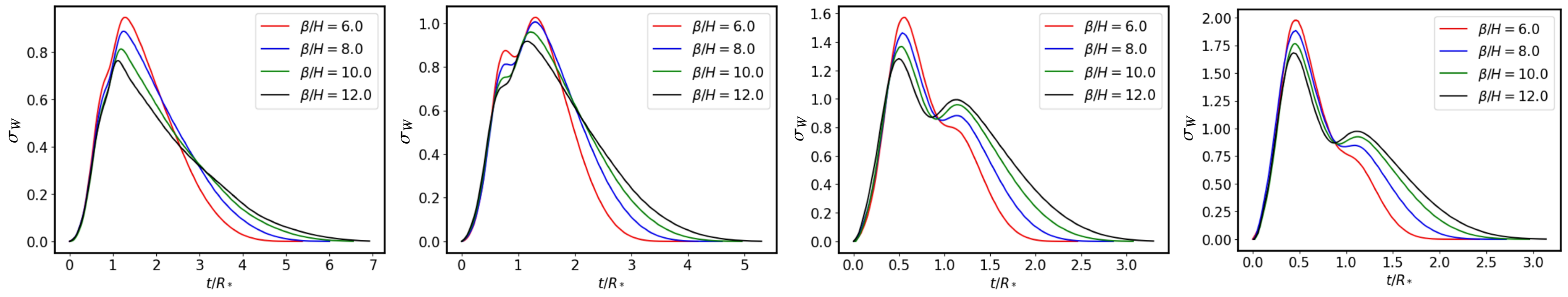}
   \caption{Evolution diagram of $\sigma_W(t)$ during FOPT process. The $\alpha$ from left to right are 0.5, 1, 5, 10, respectively. }
    \label{Fig:deltaH_all2}
\end{figure*}

Using the Press-Schechter formalism\cite{Press:1973iz,Ando:2018qdb,Yoo:2018kvb}, we obtain
\begin{equation}
    \sigma_W(t) = \int \frac{dk}{k}W^2(k) \mathcal{P_\delta}(t),
\label{Eq:delta_H_W}
\end{equation}
here, $W(k)$ is the window function, which in this case is chosen to be Gaussian
\begin{equation}
    W(k)=e^{-k^2/\mathcal{H}^2}.
\end{equation}
The results for different phase transition parameters are shown in Fig.~\ref{Fig:deltaH_all2}, which demonstrate a quite similar behavior as in Fig.~\ref{Fig:sigma_all} of the main text.

\begin{figure*}[t!]
    \centering
    \includegraphics[width=1\textwidth]{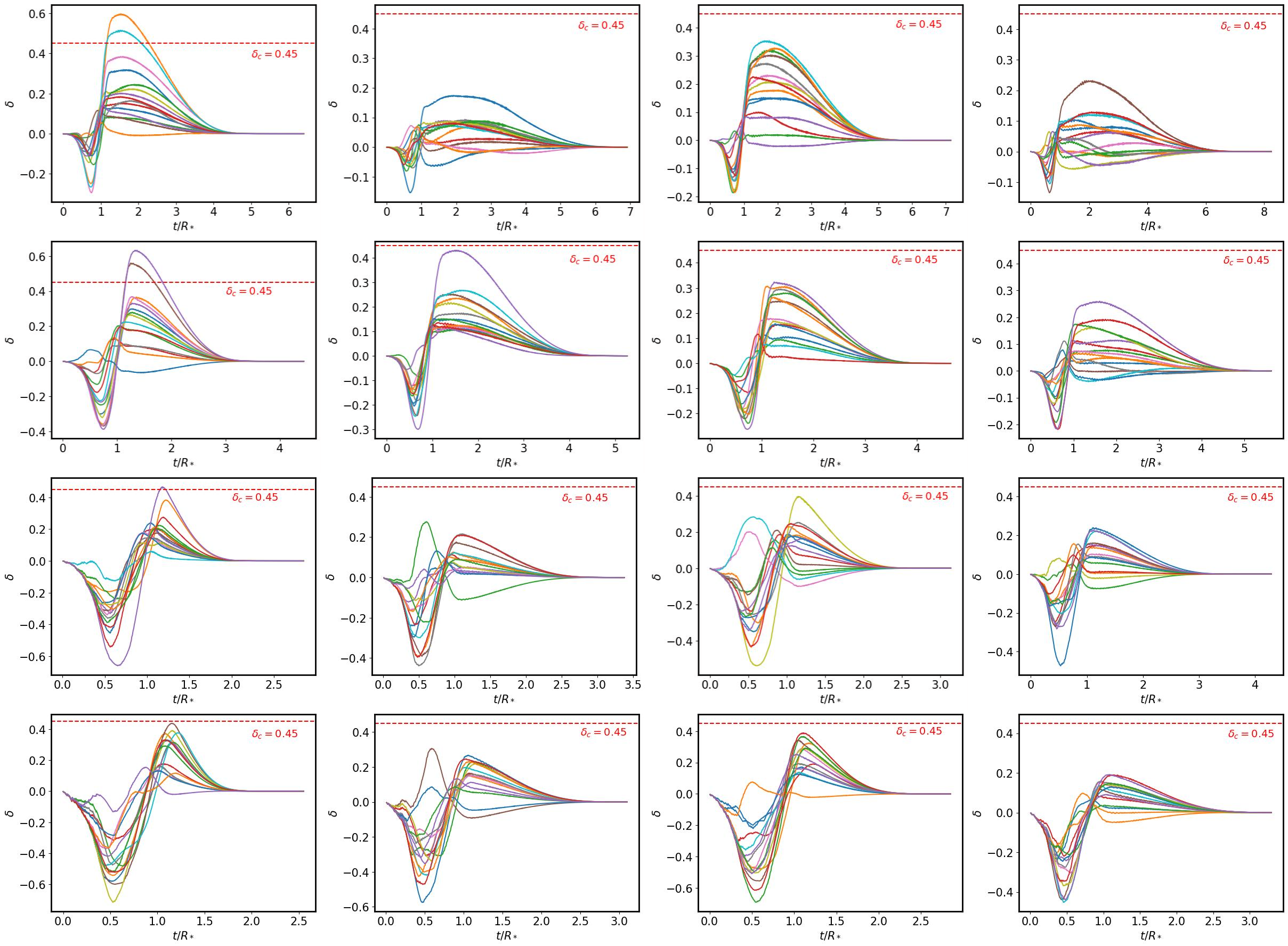}
    \caption{$\delta$ of the last 16 delayed-decayed regions. The $\beta/H$ from left to right are 6, 8, 10, and 12, the $\alpha$ from top to bottom are 0.5, 1, 5, and 10, respectively.}
    \label{delta_all}
\end{figure*}

We numerically analyze the evolution of local energy overdensities $\delta$ induced by delayed vacuum decay during FOPTs in the early universe. Specifically, we examine how different phase transition parameters, $\alpha$ and $\beta/H$, affect $\delta$ and whether these overdense regions can collapse into PBHs. As shown in the Fig.~\ref{delta_all}, due to the phase transition in the background universe, the vacuum energy gradually converts into the radiation energy, resulting in a lower energy density in the delay region compared to the background universe, and $\delta$ gradually decreases. When the delay region begins to undergo a phase transition, $\delta$ gradually increases. After the energy density disturbance reaches its maximum value, the content of the wall gradually decreases, and the background radiation energy is also averaged throughout the entire space, so $\delta$ gradually approaches zero. In addition, it can be observed that $\alpha$ has little effect on energy density perturbations, while its dependence on $\beta/H$ is extremely strong. The smaller the value of $\beta/H$, the larger the energy density perturbation. When $\beta/H$ = 6, the perturbation reaches the critical threshold $\delta_c$ = 0.45, ultimately collapsing to form PBHs.

Our simulation results show that for larger $\beta/H$ (faster phase transitions), the probability of $\delta$ exceeding the critical threshold $\delta_c=0.45$ is significantly reduced. This trend suggests that PBH formation is suppressed in faster phase transitions. This phenomenon can be understood through the following mechanisms. When $\beta/H$ is large, the phase transition proceeds rapidly, with vacuum bubbles nucleating and merging within a short timescale. This results in a more homogeneous phase transition across space, reducing the formation of localized energy overdensities.

\section{GW power spectra}
\begin{figure*}[t!]
    \centering
    \includegraphics[width=1\textwidth]{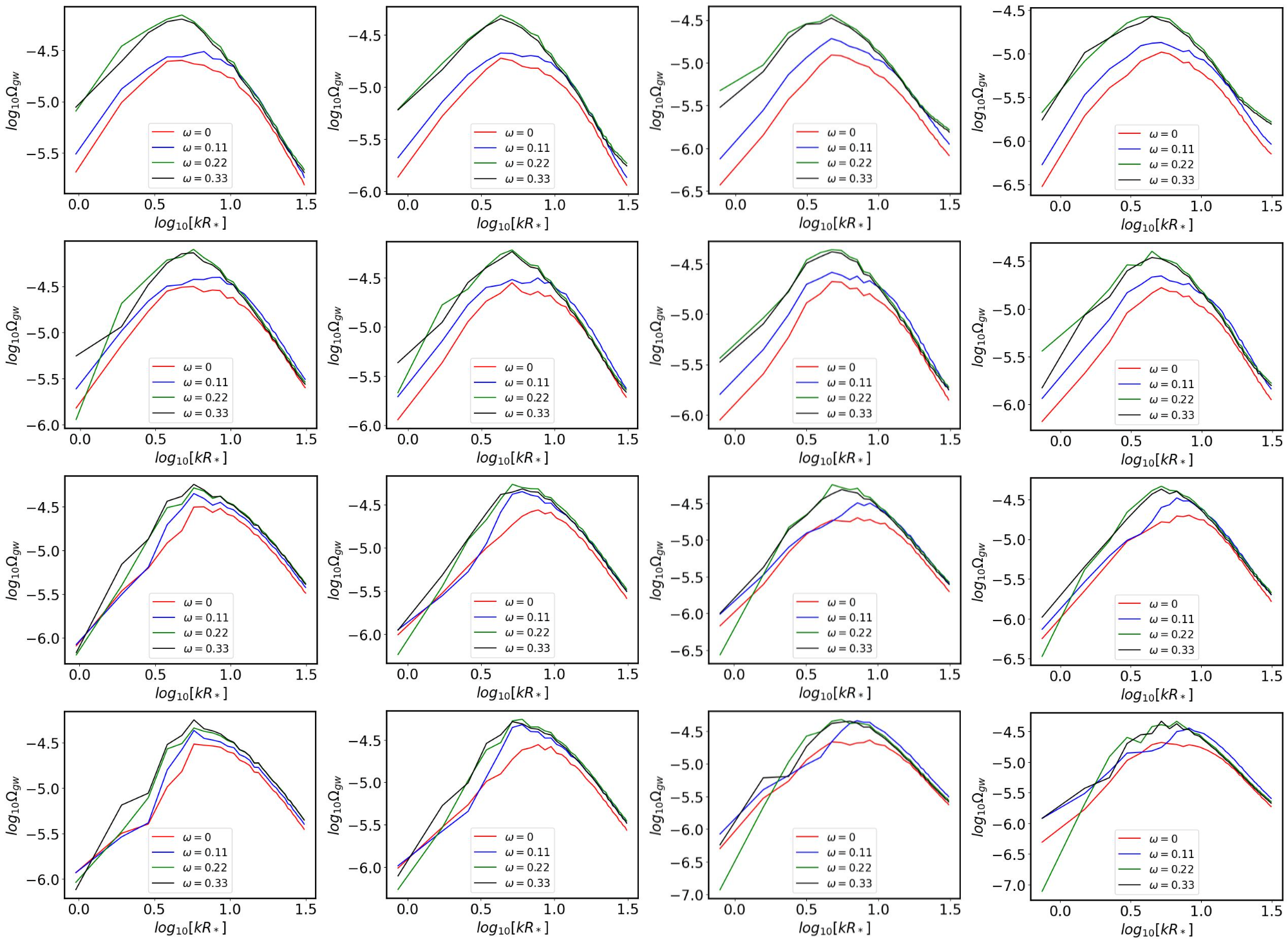}
    \caption{From left to right, the GW power spectra correspond to $\beta/H$=6, 8, 10, and 12 in sequence. The $\alpha$ from top to bottom are 0.5, 1, 5, and 10, respectively. Each subplot illustrates the energy density spectrum of GWs at the late stage of the phase transition, where the four solid lines in different colors represent the GW power spectra for different values of $\omega$.}
    \label{Fig:GW_omega}
\end{figure*}

In Fig.~\ref{Fig:GW_omega}, we compute the GW power spectrum generated during FOPTs and investigate its dependence on different values of the equation of state parameter $\omega$. Our results indicate that when $\alpha$ is 5 and 10, $\beta/H$ is 6 and 8, the slope of the GW power spectrum increases as omega increases, which do not change a lot for other circumstances. The GW power spectrum remains consistent in shape, with no significant variations observed. This suggests that the GW spectrum produced by bubble wall collisions during FOPTs is weakly dependent on $\omega$. However, we note that our conclusions regarding the dependence of the GW spectrum on $\omega$ differ from those reported in Refs.~\cite{Domenech:2020kqm,Cai:2019cdl}, requiring further investigation to clarify this issue.

\section{Fourier Analysis of a Step-Function Density Distribution}
In the ultraviolet region, the power spectrum mainly comes from contributions at very small spatial scales. In our simulations, the place where the energy density perturbation $\delta(r)$ changes most sharply on the smallest scales is near the bubble wall. There, the vacuum energy changes abruptly. This abrupt jump in energy is the main reason that determines the shape of the power spectrum in the ultraviolet region. For vacuum energy, a simple model can be defined as
\begin{equation}
    \rho(r) = \begin{cases} 0, & r < R \\ 1, & r \geq R. \end{cases}
\end{equation}
To analyze the fluctuations, we define the deviation $\delta(r)$, relative to the background
\begin{equation}
\delta(r) = \frac{\rho(r) - 1}{1} = -\Theta(R - r),
\end{equation}
where $\Theta(x)$ is the Heaviside step function. Because of spherical symmetry, we use spherical coordinates
\begin{align}
        \delta(k) &=-\int\delta(r)e^{-i\vec{k}\cdot\vec{r}}d^3r\\
        &= -\int_0^R r^2 dr \int_0^{2\pi} d\phi \int_0^\pi \sin\theta e^{-ikr \cos\theta} d\theta\\
        &=-4\pi \int_0^R r^2 \frac{\sin(kr)}{kr} dr \\
        &= -\frac{4\pi}{k} \int_0^R r \sin(kr) dr\\
        &= -\frac{R \cos(kR)}{k} + \frac{\sin(kR)}{k^2}\\
        &= -4\pi R^3 \left( \frac{\sin(kR) - kR \cos(kR)}{(kR)^3} \right).
\end{align}
At high frequencie($k \gg 1/R$),
\begin{equation}
\delta(k) \approx 4\pi R^3 \frac{kR \cos(kR)}{(kR)^3} \propto k^{-2}.
\end{equation}
Therefore, the dimensionless power spectrum follows
\begin{equation}
\mathcal{P}(k) \propto k^3 (k^{-2})^2 = k^{-1}.
\end{equation}

\end{document}